Title: In-situ formation of 2D-TiC$_x$ in Cu-Ti$_2$AlC composites: an interface reaction study

**Authors:**

***Khushbu Dash**

Postdoctoral Researcher

Dept. of Materials Engineering,

Indian Institute of Science, Bangalore

Bangalore-560012

India

Dept of Metallurgical and Materials Engg

Indian Institute of Technology, Madras

Chennai-600036

India

**Apurv Dash**

PhD scholar

Forschungszentrum Jülich GmbH,

Institute of Energy and Climate Research, 52425,

Jülich, Germany

# In-situ formation of 2D-TiC$_x$ in Cu-Ti$_2$AlC composites: an interface reaction study

K. Dash[1,2]*, A. Dash[3]

[1] Dept. of Materials Engineering, Indian Institute of Science, Bangalore, Bangalore-560012, India

[2] Dept. of Metallurgical and Materials Engineering, Indian Institute of Technology, Madras Chennai-600036, India

[3] Forschungszentrum Jülich GmbH, Institute of Energy and Climate Research: Materials Synthesis and Processing (IEK-1), 52425 Jülich, Germany

**Abstract**

In this paper, we present a concept to fabricate copper based Ti$_2$AlC MAX phase composite focusing on the processing method and the reaction which takes place at the matrix-reinforcement interface to yield 2D TiC$_x$. Copper was reinforced with Ti$_2$AlC (agglomerate size ~40 µm) phase and sintered in vacuum by pressure-less sintering. The interface of consolidated samples was investigated using transmission electron microscopy (TEM) to reveal the microstructural details. In the due course of consolidation of Cu-Ti$_2$AlC; the formation of 2D TiC$_x$ from the reaction between Cu and Ti$_2$AlC by forming solid solution between Cu and Al was facilitated. The reaction between Cu and Ti$_2$AlC has been elaborated and analyzed in the light of corroborated results. Wavelength dispersive spectroscopy (WDS) throws light on the elemental distribution at the site of interfacial reaction. This investigation elaborates a proof of concept to process an in-situ 2D TiC$_x$ reinforced Cu metal matrix composite (MMC).

Corresponding author: K. Dash

Email ID: khushbudash@gmail.com

1. Introduction

MAX phases are a class of materials which are layered ceramics, with a generic formula of $M_{n+1}AX_n$ [1]. In this formula M stands for an early transition metal, while A represents an element from the IIIA or IVA groups, and X is representative of carbon and/or nitrogen [2]. Depending on the value of n index, the MAX phases are divided to three general categories, namely 211, 312 and 413 groups. $M_{n+1}AX_n$ phases (with n = 1–3) form a wide class of nano-laminated ternary carbides or nitrides, with a hexagonal crystal structure. MAX phases have equivalent or superior electrical and thermal conductivity as compared to the corresponding metallic element M [3].

Copper based metal matrix composites (MMC) are a class of materials which combine the electrical properties of Cu and the mechanical properties of the reinforcements like ceramics particles. The second phases are very often brittle and are insulators. Hence the material suffers a compromise in the electrical properties in exchange for mechanical properties. MAX phase-based Cu MMC offers the possibility of a tough material without a profound loss in electrical conductivity. Peng et al. showed that $Ti_3AlC_2$ reinforced Cu exhibits a lower electrical conductivity but a much higher flexural strength without the loss in fracture toughness [4]. There should be no interaction between the reinforcement and the matrix for the successful fabrication of a composite. Cu is also used as an interphase to stop the reaction between Al and $Ti_3AlC_2$ for MMC applications [5]. Hence Cu is a suitable candidate for the matrix with Ti-Al based MAX phase as the reinforcement. Huang et al. demonstrated that the presence of $Ti_3AlC_2$ phase in Cu results in the increase of arc ablation resistance but other physical and mechanical properties were compromised [6]. Yu et al. extruded Mg alloy with $Ti_2AlC$ particles to induce texture and obtain anisotropic properties [7,8]. Amini et al. introduced 50 vol. % $Ti_2AlC$ resulting in the enhancement of mechanical properties of Mg [9–11]. Hu et al. reported composite of Nickel and $Ti_3AlC_2$ at 1200 °C which resulted in the decomposition of $Ti_3AlC_2$ to nano $TiC_x$ because of out-diffusion of Al. The formation of nano $TiC_x$ particles resulted in increase of flexural strength, hardness and fracture toughness [12]. Dmitruk et al. studied $Ti_3AlC_2/Ti_2AlC$-AlSi composite by squeeze casting infiltration, where the wear rate of resulting composite was less than half of Al-Si alloy alone [13]. Zhang et al. densified Cu-$Ti_3AlC_2$ composites by hot pressing at 950°C, the densification was attributed to a Cu-Al liquid phase [17]. It was shown that the tensile strength of Cu was increased to almost twice its inherent value with the addition of $Ti_3AlC_2$ whereas the tensile strength was increased to almost 3 times with the formation of in-situ Cu-$TiC_x$ composite [18]. Zhenying et al. also demonstrated the increase in flexural strength and decrease in electrical conductivity of Cu with the addition of $Ti_3AlC_2$ [19].

$(Ti_xCu_{1-x})_3(Al,Cu)C_2$ MAX phase solid solution was synthesized by sintering at 760 °C by compacting $Ti_3AlC_2$-40 vol.% Cu composite particles mixed by mechanical milling [20]. The reaction between $Ti_3AlC_2$ and Cu is elaborated by some research groups. The results show that when Cu enters the $Ti_3AlC_2$ crystal structure it yields a defective $Ti_3AlC_2$ which is $TiC_{0.67}$ with layers of Cu-Al alloy within a $Ti_3AlC_2$ grain [17]. The reports till date have emerged from high amount of $Ti_3AlC_2$ phase being added [6,17,18,21,22], but in the present work we have used only 1 and 3 vol.% of $Ti_2AlC$ phase to establish that in-situ 2D $TiC_x$ reinforced Cu composites can be fabricated with certain desirable properties by tweaking the processing conditions. Mentioning all the work done, the reaction phenomenon in case of copper and $Ti_2AlC$ has still not been explored. The properties of 2D materials are typically superior to their bulk counterparts and this motivation led to numerous researches in graphene and subsequently

MXene [14–16]. MXenes are 2D structures composed of only M and X elements and is synthesized from MAX phases by acid etching of the A element [16].

The schematic representation in Fig. 1 represents the concept behind the in-situ formation of 2D $TiC_x$ taking place in the Cu matrix reinforced with MAX phase in this experiment. The idea of this experiment was to choose a matrix which could form a solid solution with the A phase of the MAX phase. This solid solution formation would deplete the MAX phase of A element thereby creating $MX_n$ layers. The underlying principle to solid solution formation of A element with the matrix element was the eutectic reaction of Al and Cu and providing enough time for Al atoms to diffuse out of the MAX phase and react with the Cu matrix. This concept can be extended to important systems; the eutectic solid solution formation provides additional strengthening to the material. The gradient kind of microstructure distributed around with solid solution formation, 2D $TiC_x$ and Cu matrix provides ample scope to engineer the material to obtain desirable properties.

## 2. Materials and Methods

Copper powder (Alfa Aesar, 325 mesh) was blended with 1 and 3 vol. % $Ti_2AlC$ (synthesized molten salt assisted synthesis from elemental precursors [23]) agglomerate size ~40 µm) MAX phase powder using mortar and pestle followed by uniaxial cold compaction (700 MPa) and conventional sintering in vacuum furnace (Thermal Technologies, USA) at 900ºC for 2 h. The consolidated specimens were studied for microstructural characteristics and phase distribution using field emission scanning electron microscopy (FEI Quanta) and electron probe micro-analyzer (EPMA). Transmission electron microscopy (TEM) (FEI T20) has been performed to study the interaction of Cu with the $Ti_2AlC$ phase. TEM sample preparation was done by cutting a slice from the sintered disc followed by grinding to 100 µm. The 100 µm slice was punched into 3mm diameter discs which was dimpled and subsequently ion milled at 20 kV.

## 3. Results and discussion

The as-synthesized powders' $Ti_2AlC$ microstructure was investigated to show an agglomerate size of 40 micron. The X-ray diffraction pattern of starting $Ti_2AlC$ powder has been shown in Fig. 2 (a). The pattern was analyzed and shows that the powder contains $Ti_2AlC$ and traces of $Ti_3AlC_2$. Fig. 2(b) shows the starting TAC powder dispersion with an inset of higher magnification, the layers of TAC stacked together are distinct in the illustration.

### 3.1. Microstructural Analysis

The SEM microstructure and EDS analysis (Fig.2 (c & d)) shows that the grey matrix region is Cu matrix and the dark regions are $Ti_2AlC$ particles. The microstructure of the sintered specimens reveals near-uniform distribution of $Ti_2AlC$ in copper matrix. The $Ti_2AlC$ phase is observed to be layered in morphology. The EDS (Fig. 3(c) of $Ti_2AlC$ phase shows presence of Ti, Al and C along with Cu, whereas the matrix consists of Cu and Al. The core of $Ti_2AlC$ particle illustrates a different contrast as compared to periphery and Cu matrix, which suggests the occurrence of a reaction between the $Ti_2AlC$ particle and Cu matrix during sintering (Fig.3(a)). The reaction products of Cu and $Ti_2AlC$ can be seen on the $Ti_2AlC$ grains lying on the surface and the interface, EDS of spot 2 reveals eutectic formation between Cu and Al. The physical bonding of $Ti_2AlC$ in copper matrix is intimate in nature. There is appreciable physical contact between matrix and $Ti_2AlC$, and the copper from the matrix has diffused into the $Ti_2AlC$ particle which is visible from the contrast of the interface. The micrograph shows three phases by contrast which indicates Cu, Cu-Al partial solid solution and $Ti_2AlC$ particles. The phase present in the vicinity of the $Ti_2AlC$ grain is the solid solution formed by diffusion of Al and Cu atoms from the $Ti_2AlC$ particle.

The $Ti_2AlC$ phase is mostly present on the grain boundaries and pinning effect on the Cu matrix, evident by layer surrounding the grains in Fig. 3(b). Some twins are visible in the Cu matrix which perhaps formed during sintering of the composite, also increases the mechanical performance of the composite. The $Ti_2AlC$ phase embedded in the copper matrix Fig. 4(c) suggests that Cu has reacted and penetrated into the edge containing $TiC_x$ layer thereby forming Cu-Al alloy, dissolving Al partially in due process [22]. An indication of 2D $TiC_x$ formation is the sheet like loose ends at the grain boundary regions labelled by arrows in Fig. 4(b). The $Ti_2AlC$ phase is present as layered particle along with sheet like features which suggests that 2D $TiC_x$ has been formed in certain regions with existing $Ti_2AlC$ phase. Zhang et al. [17] suggested that Cu and $Ti_3AlC_2$ reacts above 850°C to form a layer at the interface. Molten Cu triggers the exfoliation of $Ti_3AlC_2$ via its decomposition yielding $TiC_{0.67}$ layers and Cu-Al alloy layers within $Ti_3AlC_2$ grains [18].

### 3.2. Elemental mapping at the interface

The EPMA-WDS analysis (Fig. 4) shows that the Al in the $Ti_2AlC$ phase has interacted with the Cu matrix to form a partial solid solution. The WDS of the $Ti_2AlC$ phase particle, the interfacial area and the matrix has been illustrated in Fig. 4. The interphase consists of Al in an appreciable amount. The core of the particle shows lower concentration (atomic %) of Al in accordance to the Al content in $Ti_2AlC$ phase, as it has interacted with copper and has diffused into the matrix region. The periphery of the interphase region is Al rich indicating the onset of formation of a Cu-Al solid solution.

It can be observed that Cu has entered the core of $Ti_2AlC$ phase which proves the reaction taking place between Cu and $Ti_2AlC$ phase result in a partial Cu-Al solid solution. The outward diffusion of Al from $Ti_2AlC$ to the Cu matrix is visible from EPMA maps, the Al map shows the presence of Al in the Cu matrix. This confirms that the Al diffuses into the Cu matrix forming a partial solid solution with Cu. The Cu map indicates the inward diffusion of Cu from the matrix into the $Ti_2AlC$ particle substituting in the Ti atomic position as well [20]. The presence of Cu alters the decomposition temperature of $Ti_2AlC$, i.e. generally the stability of $Ti_2AlC$ ranges from 1100 to 1200 °C, but Cu decreases the stability temperature considerably by reacting with $Ti_2AlC$ at 900 °C [17]. These diffusion reactions can be formulated as

$$Ti_2AlC + Cu \longrightarrow TiC_x + CuAl_x$$

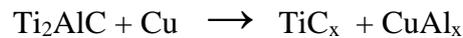

Cu and $Ti_2AlC$ have been reported to have a good wettability with each other, which justifies the Cu entering into the core of $Ti_2AlC$ by capillary action and diffusion. The de-intercalation of Al in MAX phase is most viable as the M-A bond is quite weak. The decomposition of MAX phase is dependent on temperature. A mild reaction between Cu and $Ti_2AlC$ starts at 850 -950 °C, the reaction becomes strong above 950 °C. Cu-Al solid solution formation has been studied when $Ti_3AlC_2$ was reinforced in Cu matrix [17]. EPMA analysis reveals the presence of Cu in the core of the TAC particles, due to which the sub layers are visible. At temperatures above than 1000 °C, total decomposition occurs with sub-stoichiometric $TiC_x$ and Cu-Al solid solution. Earlier research showed that Cu enters $Ti_3AlC_2$ phase but location of Cu was not specified. The products from the diffusion can be summarized as showed in the schematic Fig. 1 which specifies formation of 2D $TiC_x$; solid solution of Cu-Al, remaining Cu and some defective MAX phase.

### 3.3. Reaction at the interface

Microstructure of the Cu- $Ti_2AlC$ composite reveals dislocation tangles in the Cu matrix (Fig. 5(a)); as the $Ti_2AlC$ phase is layered, the whole structure has imbibed the layered alignment; showing dislocation arrays. These arrays may be a result of consolidation which induces some deformation of matrix and reinforcement. The TEM microstructures show a layered structure of $Ti_2AlC$ phase in the copper matrix shown in Fig. 5 (b). The TEM-EDS results reveal that the dark region is $Ti_2AlC$ phase and the lighter area is copper matrix (Fig. 5(c)). The SAD patterns show continuous streaks which prove that layered structure is present. The bright field micrographs in Fig. 5 (d) shows two distinct regions of interest. i.e. near the Cu- $Ti_2AlC$ phase interface and away from the interface. The region near the interface exhibits a sub-grain boundary pattern which is spread over the area which is close to the interface. This area can be marked as the reaction layer where Cu and $Ti_2AlC$ have interacted to form 2D $TiC_x$, in due course Cu has entered the $Ti_2AlC$ phase and formed partial solid solution of Cu-Al (Fig. 5 (d)). The region farther from the interface shows some precipitates which can be identified by viewing the bright and dark field images. These precipitates denote the formation of solid solution between Cu-Al. Cu was found within the defective $Ti_3AlC_2$ and a crystallographic relationships ½1210 $Ti_3AlC_2$ ½111Cu and (0001) $Ti_3AlC_2$ (111) Cu was established [17].

Moire pattern visible in Fig. 5 (e) which form due to interdiffusion of Cu and $TiC_x$ phase, as stated by Zhang et al. [18] showing circular and parallel patterns. The 2D $TiC_x$ possess a defective crystal structure; hence the superimposition of defective crystals gives rise to Moire fringes. Some single sheets of 2D $TiC_x$ are visibly present in the matrix from Fig. 5(f &g) which clearly shows electronically transparent area of $TiC_x$. The $TiC_x$ sheets are of 50 nm in average width. The $TiC_x$ nanosheets (monolayers) can be seen at the grain boundaries of Cu matrix indicating the presence of $Ti_2AlC$ phase at the interface, hence the reaction and formation of 2D $TiC_x$ which is the reaction site between $Ti_2AlC$ and Cu as well matrix strengthening location. TEM microstructures corroborates the above fact of two dimensional entities being present at the grain boundaries. Single 2D $TiC_x$ flakes are also visible in abundant quantity ; whereas the SAD pattern of these transparent flakes shows (002) ring present in them (Fig.5(h)). $TiC_x$ flakes are fragmented during the diffusion reaction process mentioned above, hence show size in nanometer scale. The formation of $TiC_x$ is anticipated to have been taken place partially i.e. localized in the periphery of $Ti_2AlC$ phase regions. We can see black region in some areas which is confirmed as the $Ti_2AlC$ phase by EDS where as we also observe transparent entities. This fact could be attributed to the partial solid solution formation between Cu from the matrix and Al from the $Ti_2AlC$ phase yielding 2D $TiC_x$ which can be observed in the specimen. As the matrix reveals the presence of Cu, Al as well the $Ti_2AlC$, the SEM combined with TEM microstructures propose that here has been some interaction between Cu and Al that has led to the formation of partial solid solution during the sintering of the composite.

## 4. Conclusions

The following points can be concluded from the exploration:

1. This design of experimentation was performed for the first-time using powder metallurgy process, yielding in-situ 2D $TiC_x$ in the Cu matrix with partial Cu-Al solid solution by processing Cu and $Ti_2AlC$ together.

2. The microstructure of the composites were explained which showed a typical characteristic of sub-layers of the MAX phase in the copper matrix certifying the diffusion of Cu and Al atoms in and out of $Ti_2AlC$.

3. The reaction between Cu and $Ti_2AlC$ is studied by TEM as well as EPMA-WDS which shows that Cu diffuses into the $Ti_2AlC$ and forms a partial solid solution of Cu-Al and forms 2D $TiC_x$.

4. TEM reveals the formation of 2D $TiC_x$ at certain sites preferably at grain boundaries, due to low amount of $Ti_2AlC$ added to the composite. Formation of single flakes of 2D $TiC_x$ has been shown in the Cu matrix at the grain boundaries.

**Acknowledgement**

The authors would like to thank the infrastructural support from Indian Institute of Science, Bangalore and Forschungszentrum Jülich GmbH. K.D. would like to acknowledge the SERB NPDF grant no: PDF/2016/003051 and the microscopy facilities at Indian Institute of Science, Bangalore.

**References**

[1] M.W. Barsoum, The MN+1AXN phases: A new class of solids: Thermodynamically stable nanolaminates, Prog. Solid State Chem. 28 (2000) 201–281. doi:10.1016/S0079-6786(00)00006-6.

[2] M.W. Barsoum, MAX Phases: Properties of Machinable Ternary Carbides and

Nitrides, Wiley-VCH Verlag GmbH & Co. KGaA, Weinheim, Germany, 2013. doi:10.1002/9783527654581.

[3]     M.W. Barsoum, T. El-Raghy, Synthesis and characterization of a remarkable ceramic: Ti3SiC2, J. Am. Ceram. Soc. 79 (1996) 1953–1956. doi:10.1111/j.1151-2916.1996.tb08018.x.

[4]     L. Peng, Fabrication and properties of Ti3AlC2 particulates reinforced copper composites, Scr. Mater. 56 (2007) 729–732. doi:10.1016/j.scriptamat.2007.01.027.

[5]     S. Wang, S. Zhu, J. Cheng, Z. Qiao, J. Yang, W. Liu, Microstructural, mechanical and tribological properties of Al matrix composites reinforced with Cu coated Ti3AlC2, J. Alloys Compd. 690 (2017) 612–620. doi:10.1016/J.JALLCOM.2016.08.175.

[6]     X. Huang, Y. Feng, G. Qian, H. Zhao, J. Zhang, X. Zhang, Physical, mechanical, and ablation properties of Cu–Ti 3 AlC 2 composites with various Ti 3 AlC 2 contents, Mater. Sci. Technol. 34 (2018) 757–762. doi:10.1080/02670836.2017.1407900.

[7]     W. Yu, H. Zhao, X. Hu, Anisotropic mechanical and physical properties in textured Ti2AlC reinforced AZ91D magnesium composite, J. Alloys Compd. 732 (2018) 894–901. doi:10.1016/j.jallcom.2017.10.255.

[8]     W. Yu, H. Zhao, X. Wang, L. Wang, S. Xiong, Z. Huang, S. Li, Y. Zhou, H. Zhai, Synthesis and characterization of textured Ti2AlC reinforced magnesium composite, J. Alloys Compd. 730 (2018) 191–195. doi:10.1016/j.jallcom.2017.09.303.

[9]     S. Amini, C. Ni, M.W. Barsoum, Processing, microstructural characterization and mechanical properties of a Ti2AlC/nanocrystalline Mg-matrix composite, Compos. Sci. Technol. 69 (2009) 414–420. doi:10.1016/j.compscitech.2008.11.007.

[10]    S. Amini, M.W. Barsoum, On the effect of texture on the mechanical and damping properties of nanocrystalline Mg-matrix composites reinforced with MAX phases, Mater. Sci. Eng. A. 527 (2010) 3707–3718. doi:10.1016/j.msea.2010.01.073.

[11] B. Anasori, S. Amini, V. Presser, M.W. Barsoum, Nanocrystalline Mg-Matrix Composites with Ultrahigh Damping Properties, in: Magnes. Technol. 2011, John Wiley & Sons, Inc., Hoboken, NJ, USA, 2011: pp. 463–468. doi:10.1002/9781118062029.ch87.

[12] W. Hu, Z. Huang, L. Cai, C. Lei, H. Zhai, S. Hao, W. Yu, Y. Zhou, Preparation and mechanical properties of TiCx-Ni3(Al,Ti)/Ni composites synthesized from Ni alloy and Ti3AlC2 powders, Mater. Sci. Eng. A. 697 (2017) 48–54. doi:10.1016/J.MSEA.2017.04.113.

[13] A. Dmitruk, A. Żak, K. Naplocha, W. Dudziński, J. Morgiel, Development of pore-free Ti-Al-C MAX/Al-Si MMC composite materials manufactured by squeeze casting infiltration, Mater. Charact. 146 (2018) 182–188. doi:10.1016/j.matchar.2018.10.005.

[14] M. Ghidiu, M.R. Lukatskaya, M. Zhao, Y. Gogotsi, M.W. Barsoum, Conductive two-dimensional titanium carbide 'clay' with high volumetric capacitance, Nature. 516 (2014) 78–81. doi:10.1038/nature13970.

[15] B. Anasori, M.R. Lukatskaya, Y. Gogotsi, 2D metal carbides and nitrides (MXenes) for energy storage, Nat. Rev. Mater. 2 (2017) 16098. doi:10.1038/natrevmats.2016.98.

[16] M. Naguib, M. Kurtoglu, V. Presser, J. Lu, J. Niu, M. Heon, L. Hultman, Y. Gogotsi, M.W. Barsoum, Two-dimensional nanocrystals produced by exfoliation of Ti3AlC2, Adv. Mater. 23 (2011) 4248–4253. doi:10.1002/adma.201102306.

[17] J. Zhang, J.Y. Wang, Y.C. Zhou, Structure stability of Ti3AlC2 in Cu and microstructure evolution of Cu-Ti3AlC2 composites, Acta Mater. 55 (2007) 4381–4390. doi:10.1016/j.actamat.2007.03.033.

[18] J. Zhang, Y.C. Zhou, Microstructure, mechanical, and electrical properties of Cu-Ti3AlC2 and in situ Cu-TiCx composites, J. Mater. Res. 23 (2008) 924–932. doi:10.1557/jmr.2008.0126.

[19] Z. Huang, H. Zhai, M. Li, X. Chen, High Performance of Sub-Micro-Layered Ti 3 C 2 /(Cu-Al) Cermets Prepared by In-Situ Hot-Extruding Method, (2010). doi:10.4028/www.scientific.net/MSF.654-656.2049.

[20] M. Nechiche, V. Mauchamp, A. Joulain, T. Cabioc, X. Milhet, P. Chartier, S. Dubois, Synthesis and characterization of a new (Ti1-ε,Cuε)3(Al,Cu)C2 MAX phase solid solution, J. Eur. Ceram. Soc. 3 (2016) 459–466. doi:10.1016/j.jeurceramsoc.2016.09.028.

[21] M. Ai, H. Zhai, Z. Tang, Interformational exfoliation of Ti3AlC2 induced by Cu, 2007. doi:10.4028/www.scientific.net/KEM.336-338.1371.

[22] Z. Huang, J. Bonneville, H. Zhai, V. Gauthier-Brunet, S. Dubois, Microstructural characterization and compression properties of TiC0.61/Cu(Al) composite synthesized from Cu and Ti3AlC2 powders, J. Alloys Compd. 602 (2014) 53–57. doi:10.1016/J.JALLCOM.2014.02.159.

[23] A. Dash, R. Vaßen, O. Guillon, J. Gonzalez-Julian, Molten salt shielded synthesis of oxidation prone materials in air, Nat. Mater. 18 (2019) 465–470. doi:10.1038/s41563-019-0328-1.

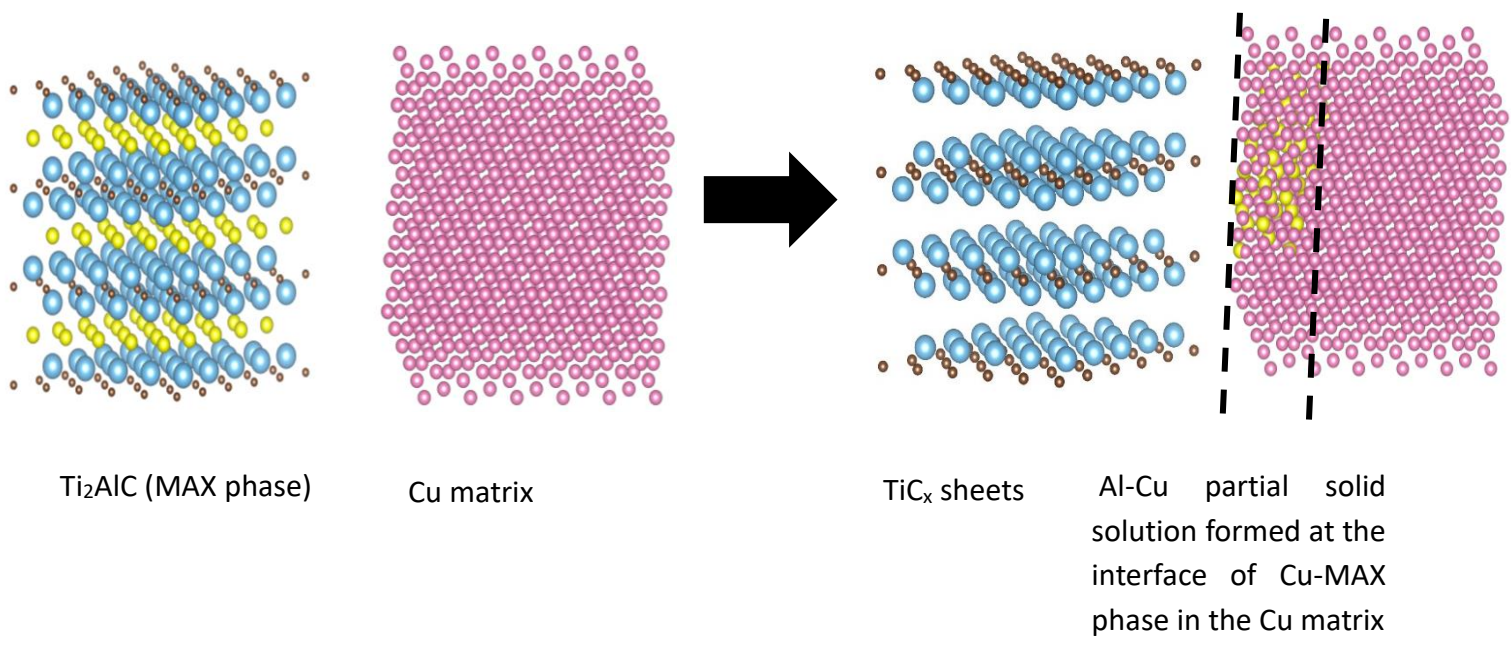

Fig. 1 Schematic representation of the in-situ reaction taking place in the process of sintering

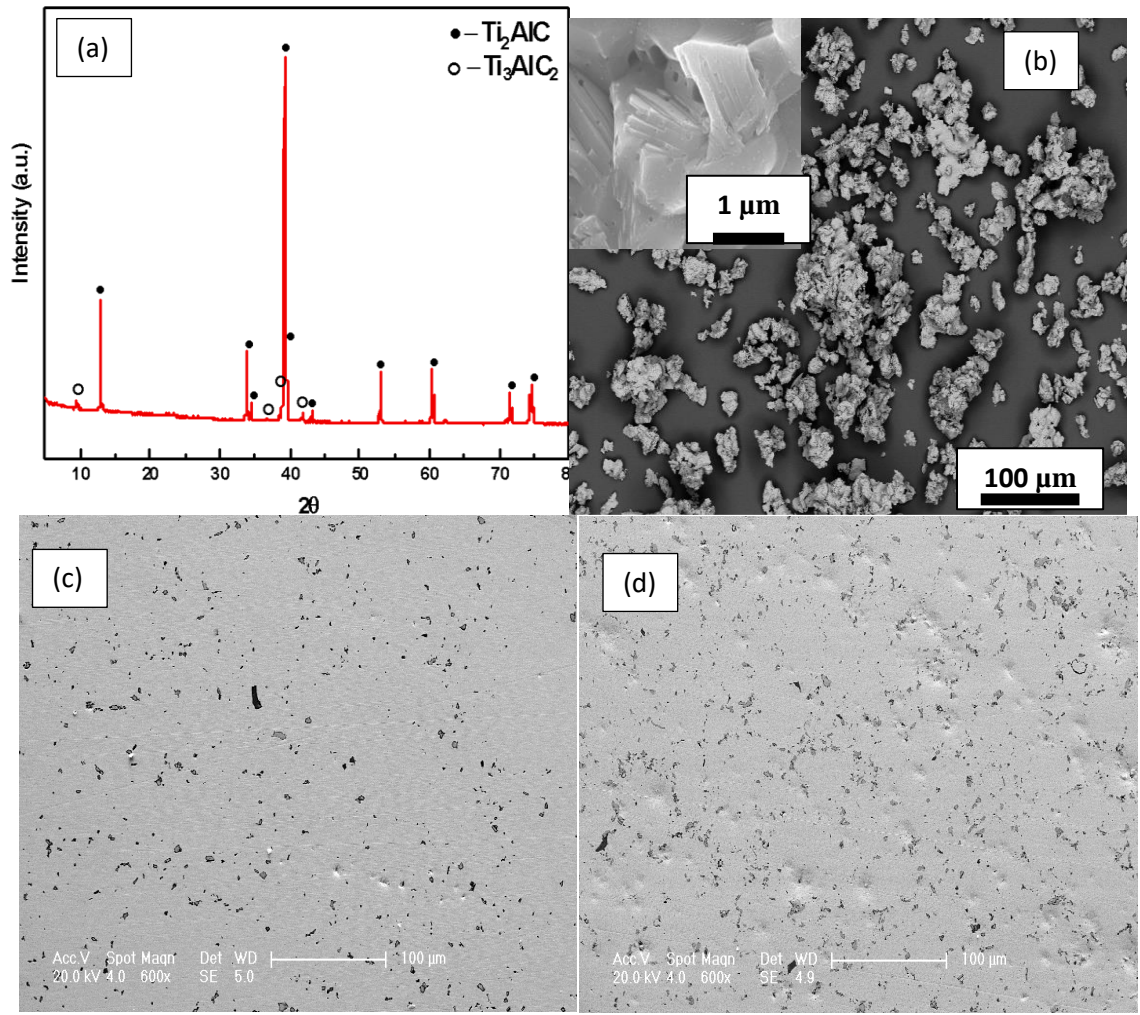

Fig. 2 (a) Phase analysis of starting Ti$_2$AlC powder used as reinforcement (b) Microstructure of Ti$_2$AlC powder, (c&d) microstructure of Cu-1&3 vol.% Ti$_2$AlC composite showing distribution of particles in matrix.

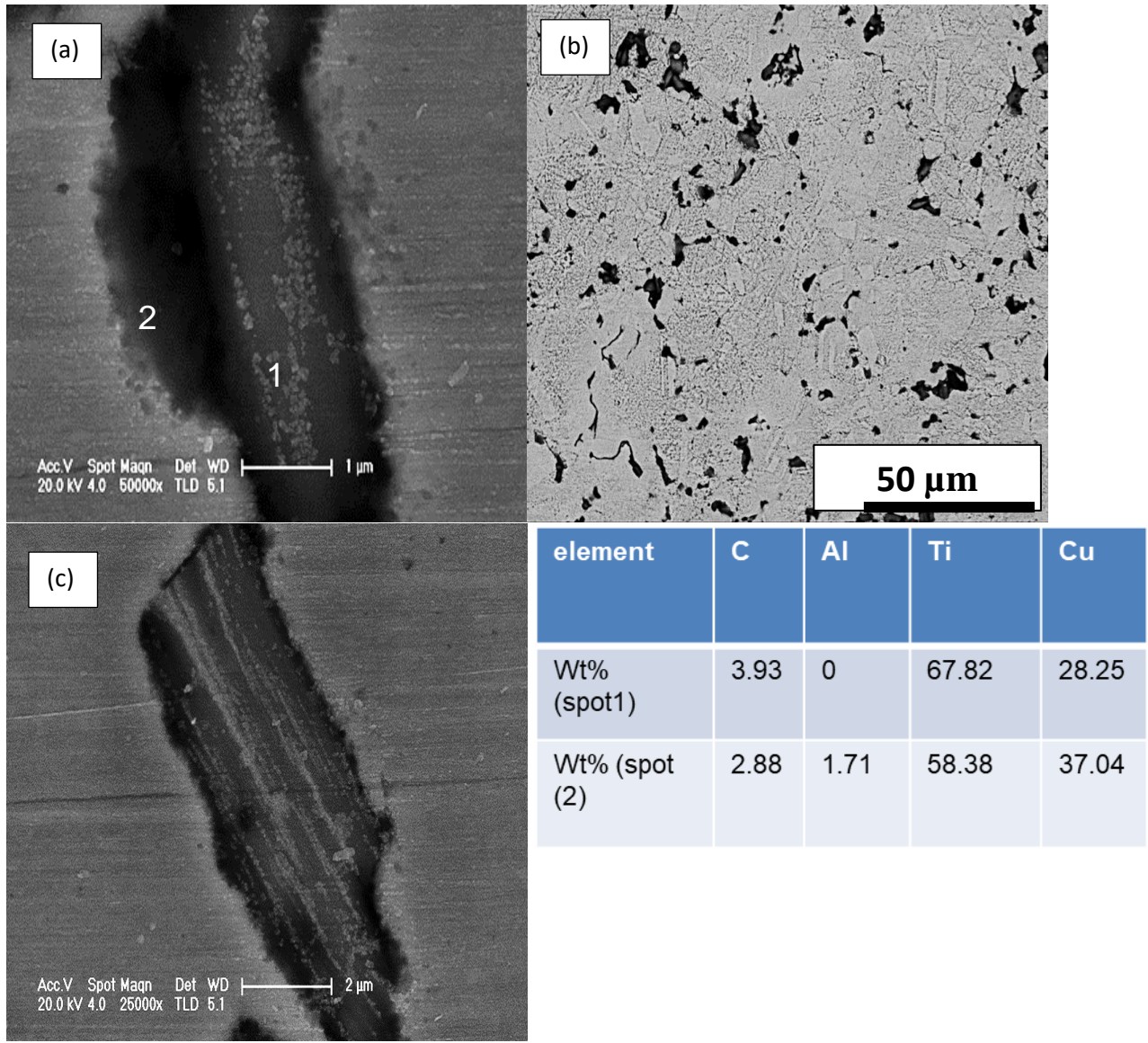

Fig. 3 (a) Microstructure of Cu-3 vol.% Ti$_2$AlC composite at high magnification and corresponding EDS analysis showing the elemental analysis of diffused region at the interface and the core of the Ti$_2$AlC particle; (b) Microstructure showing 3 contrast regions (c) SEM micrograph illustrating sub layers in Ti$_2$AlC phase particle embedded in Cu matrix.

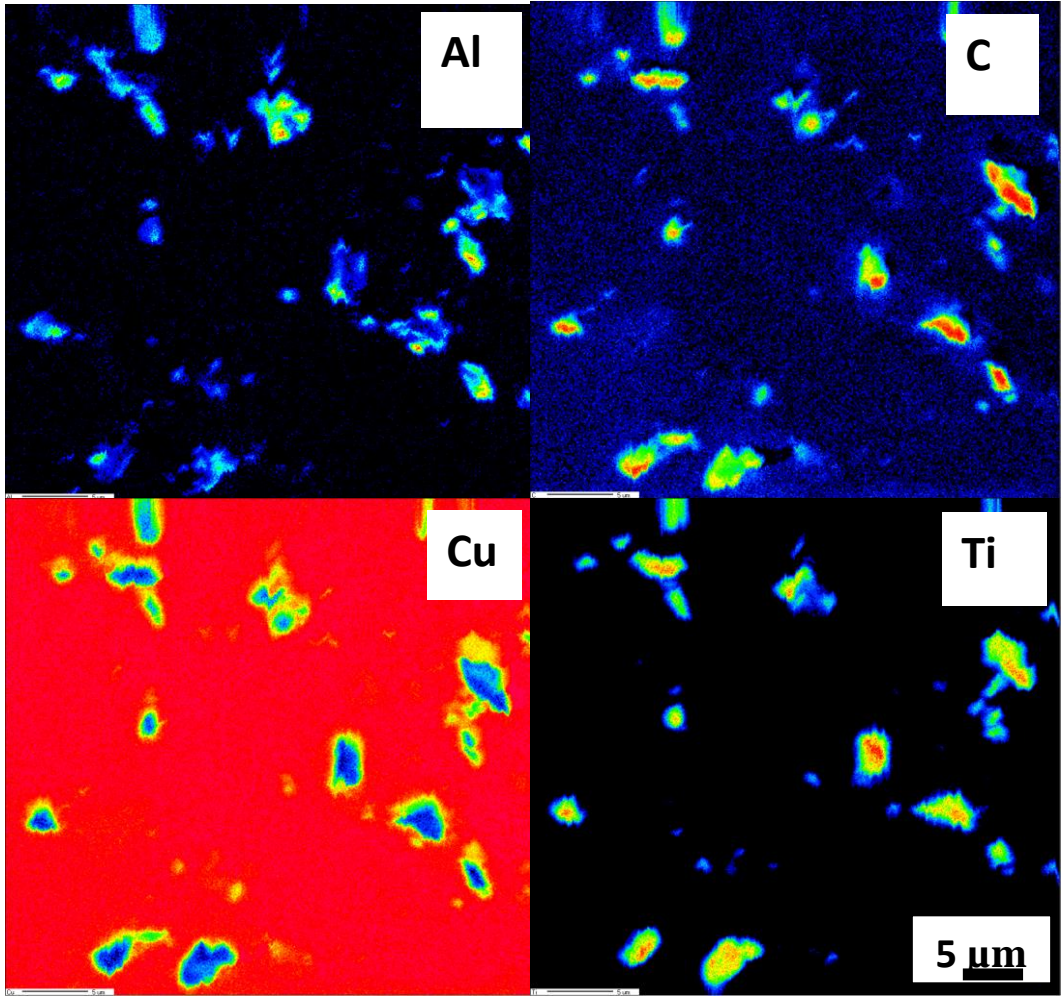

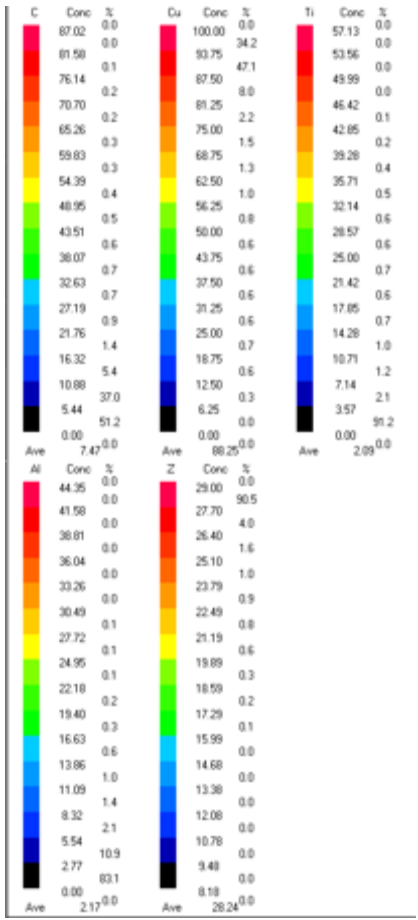

Figure 4. Elemental mapping of (a) Al, (b) C, (c) Cu and (d) Ti in Cu-3 vol.% Ti$_2$AlC metal matrix composite

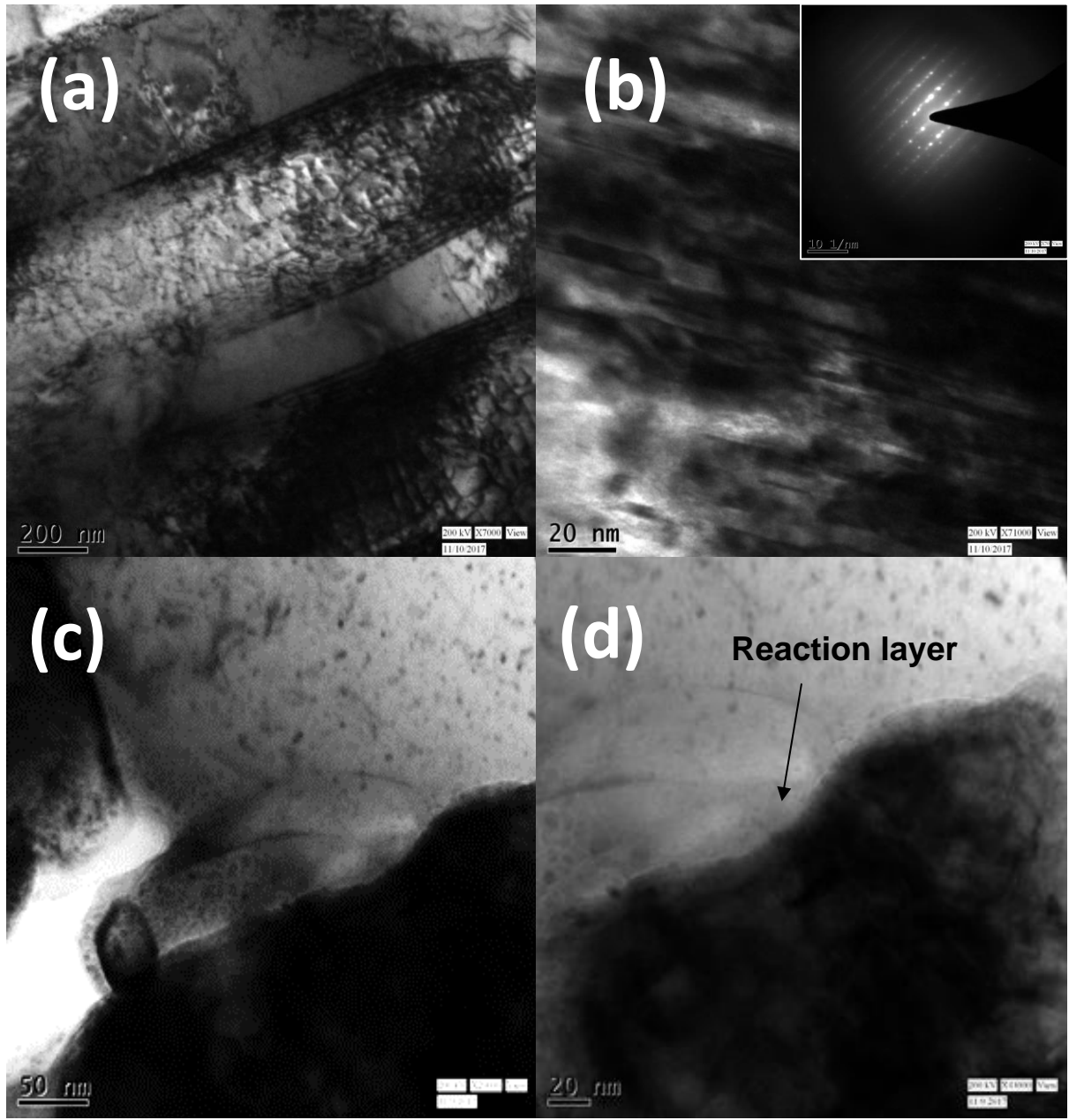

Fig.5 TEM micrographs of Cu-3 vol.% Ti$_2$AlC composite illustrating (a) dislocation in layers (b) layered stacked together (inset corresponding SAD pattern) (c) interaction between Cu and Ti$_2$AlC phase and (d) reaction layer between Cu and Ti$_2$AlC phase

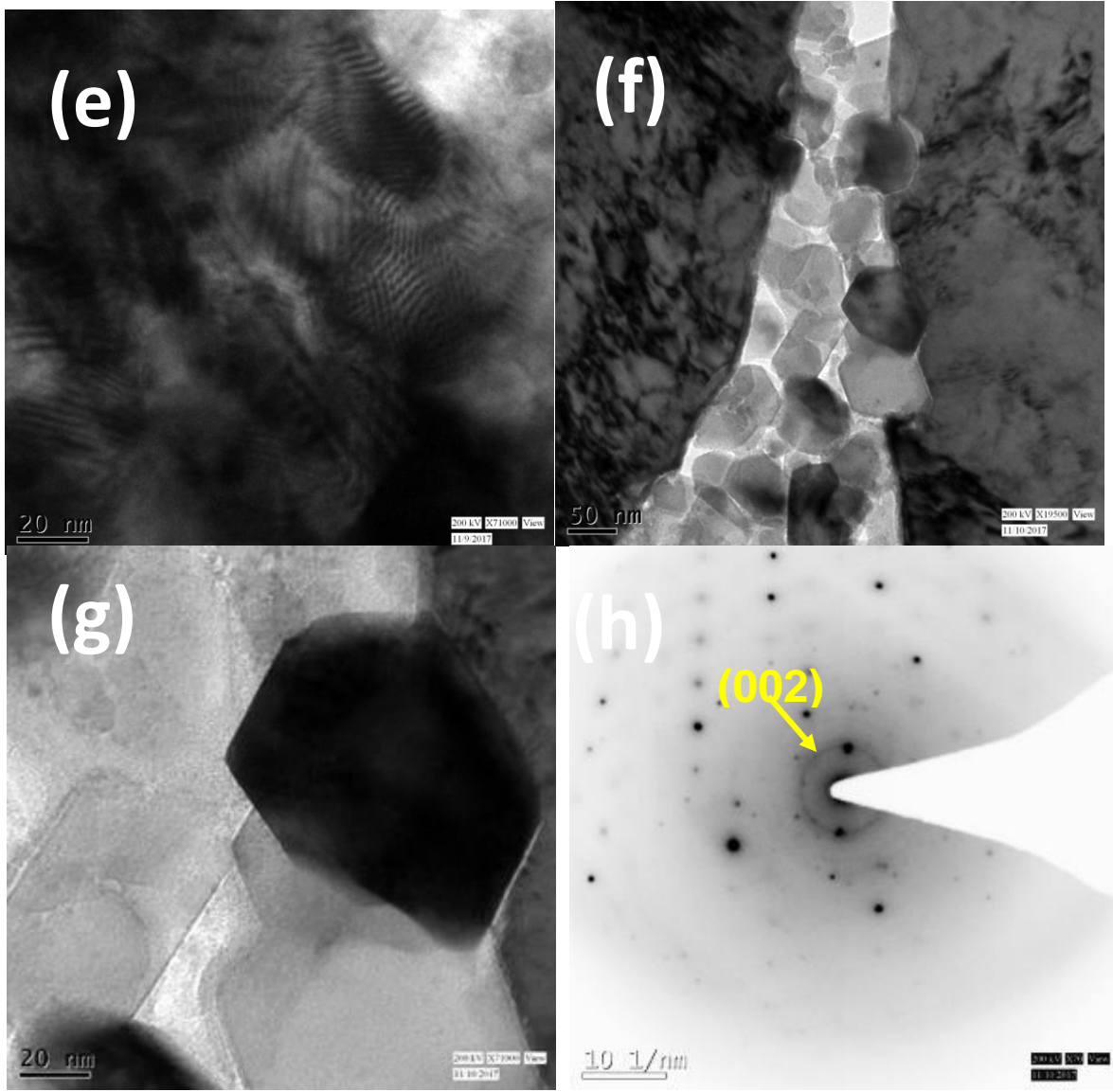

Fig. 5 TEM micrographs of Cu-3 vol.% $Ti_2AlC$ composite showing (e) moire fringes in the Cu-$Ti_2AlC$ composite (f) 2D $TiC_x$ at the grain boundary (g) magnified single flakes of 2D $TiC_x$ (h) SAD pattern of 2D $TiC_x$